\documentclass[a4paper,11pt]{article}
\usepackage[utf8]{inputenc}
\usepackage{amsmath}
\usepackage{amsfonts}
\usepackage{amssymb}
\usepackage{amsthm}
\usepackage{caption}
\usepackage{multirow}
\usepackage{fontenc}
\usepackage{graphicx}
\usepackage{color}
\usepackage{hyperref}
\usepackage[symbol]{footmisc}
\usepackage[total={6in, 9.5in}]{geometry}
\usepackage{subcaption}
\date{}

\def\title#1{\begin{center} {\LARGE #1 \vspace{0.5cm}} \end{center}}
\def\author#1{\begin{center}{ \large #1} \end{center}}
\def\affil#1{\begin{center}{ \it #1} \end{center}}
\def\email#1{\begin{center}{\normalsize #1} \end{center}}
\def\preprint#1{\rightline{\begin{tabular}{l} #1 \end{tabular}}}
\def\acknowledgement#1{{\noindent \bf\large Acknowledgements} \,\,#1 }
\begin{document}
\preprint{ABCD-ABCD} 

\title{Baby MIND detector first physics run} 

\def\author#1{\begin{center}{ \large #1} \end{center}}
\def\affil#1{\begin{center}{ \it #1} \end{center}}

\small{
\author{ \small A.~ Ajmi$^{1}$ ,Y.~ Asada$^{2}$, P.~Benoit$^3$, A.~Blondel$^4$, M.~Bogomilov$^5$, A.~ Bonnemaison$^{6}$, A.~Bross$^7$, F.~Cadoux$^4$, S.~ Cao$^{8}$, A.~Cervera$^{9}$, N.~Chikuma$^{10}$, R.~ Cornat$^{6}$, L.~ Domine$^{6}$, O.~ Drapier$^{6}$, A.~Dudarev$^3$, A.~ Eguchi$^{10}$, T.~Ekel\"of$^{11}$, Y.~Favre$^4$, S.~Fedotov$^{12}$, O.~ Ferreira$^{6}$, F.~ Gastaldi$^{6}$, M.~ Gonin$^{6}$,   Y.~ Hayato$^{13}$, A.~ Hiramoto$^{1}$, T.~ Honjo$^{14}$, A.K.~ Ichikawa$^{1}$, J.~ Imber$^{6}$, M.~Khabibullin$^{12}$, A.~Khotyantsev$^{12}$, T.~ Kikawa$^1$, K.~ Kin$^{14}$,T.~ Kobata$^{14}$ ,T.~ Kobayashi$^{8}$,  A.~Kostin$^{12}$, Y.~Kudenko$^{12}$, N.~ Kukita$^{14}$, S.~ Kuribayashi$^{1}$, M.~ Louzir$^{6}$  G.~Mitev$^{15}$, K.~ Matsushita$^{1}$, A.~Mefodiev$^{12}$, A.~Minamino$^{2}$,  F.~ Magniette$^{6}$, O.~Mineev$^{12}$, T.~ Mueller$^{6}$, T. Nakaya$^{1}$, M.~Nessi$^3$, L.~Nicola$^4$, K.~ Nishiziki$^{14}$, E.~Noah$^4$, J.~Nugent$^{16}$, T.Odagawa$^{1}$, K.~ Okamoto$^{2}$, H.~Pais Da Silva$^3$, S.~Parsa $^{4 \, \dagger}$ \footnote[5]{Speaker}\,\footnote[1]{On behalf of WAGASCI - Baby MIND collaboration}, G.~	Pintaudi$^{2}$, B.~Quilain $^{10}$, M.~ Rayner$^{3}$, G.~Rolando$^3$, C.~ Ruggles$^{16}$ F.~Sanchez$^4$, Y.~ Seiya$^{14}$,  F.J.P.~Soler$^{16}$,  S.~Suvorov$^{12}$, S.Takayasu$^{14}$, S.~ Tanaka$^{14}$, H.~Ten Kate$^3$, N.~ Teshima$^{14}$, N.~ Tran$^{17}$, R.~Tsenov$^5$,  G.~Vankova-Kirilova$^5$, L.~ Vignoli$^{6}$, O.~ Volcy$^{6}$,  K. Yamamoto$^{14}$, K.~ Yasutome$^{1}$, N.~Yershov$^{12}$ and  M.~ Yokoyama$^{10}$.}

\affil{$^{1}$Kyoto University, Kyoto, Japan.\\
$^{2}$Yokohama National University, Yokohama, Japan.\\
$^3$European Organization for Nuclear Research, CERN, Geneva, Switzerland.\\
$^4$University of Geneva, Section de Physique, DPNC, Geneva, Switzerland.\\
$^5$University of Sofia, Department of Physics, Sofia, Bulgaria.\\
$^{6}$Ecole Polytechnique, IN2P3-CNRS, Laboratoire Leprince-Ringuet, Palaiseau, France\\
$^7$Fermi National Accelerator Laboratory, Batavia, Illinois, USA.\\
$^{8}$High Energy Accelerator Research Organization (KEK), Tsukuba, Ibaraki, Japan\\
$^{9}$IFIC (CSIC $\&$ University of Valencia), Valencia, Spain.\\
$^{10}$University of Tokyo, Tokyo, Japan.\\
$^{11}$University of Uppsala, Uppsala, Sweden.\\
$^{12}$Institute of Nuclear Research, Russian Academy of Sciences, Moscow, Russia.\\
$^{13}$University of Tokyo, Institute for Cosmic Ray Research, Kamioka Observatory, Kamioka, Japan\\
$^{14}$Osaka City University, Department of Physics, Osaka, Japan\\
$^{15}$Academy of Sciences, Sofia, Bulgaria\\
$^{16}$University of Glasgow, School of Physics and Astronomy, Glasgow, UK.\\
$^{17}$IFRISE, Quy Nhon, Vietnam.
}
}

\def\author#1{\begin{center}{ \sc #1} \end{center}}
\def\affil#1{\begin{center}{ \it #1} \end{center}}

%

\email{$^{\dagger}$saba.parsa@unige.ch}
%
%
%

\begin{abstract}
Baby MIND is a Magnetized Iron Neutrino Detector, serving as a downstream magnetized muon range detector for WAGASCI on the T2K beam line in Japan.
The first physics run of Baby MIND together with other WAGASCI sub-detectors took place in the period from November 2019 to February 2020 (T2K run10), where a total of $4.8 \times 10^{20}$ Protons on target (POT) was delivered. Preliminary results showing Baby MIND data quality, detector performance and examples of neutrino interactions on iron during the first physics run are presented. 
\end{abstract}
\vfill
\begin{center}
{\LARGE{Presented at}}\\
\vspace{1cm}
\Large{NuPhys2019: Prospects in Neutrino Physics\\
Cavendish Conference Centre, London, 16--18 December 2019}
\end{center}

\clearpage

 
\section{Introduction}
\begin{figure}[h!]
\centering
\begin{minipage}{0.5\textwidth}
	\centering
	\includegraphics[height = 4cm]{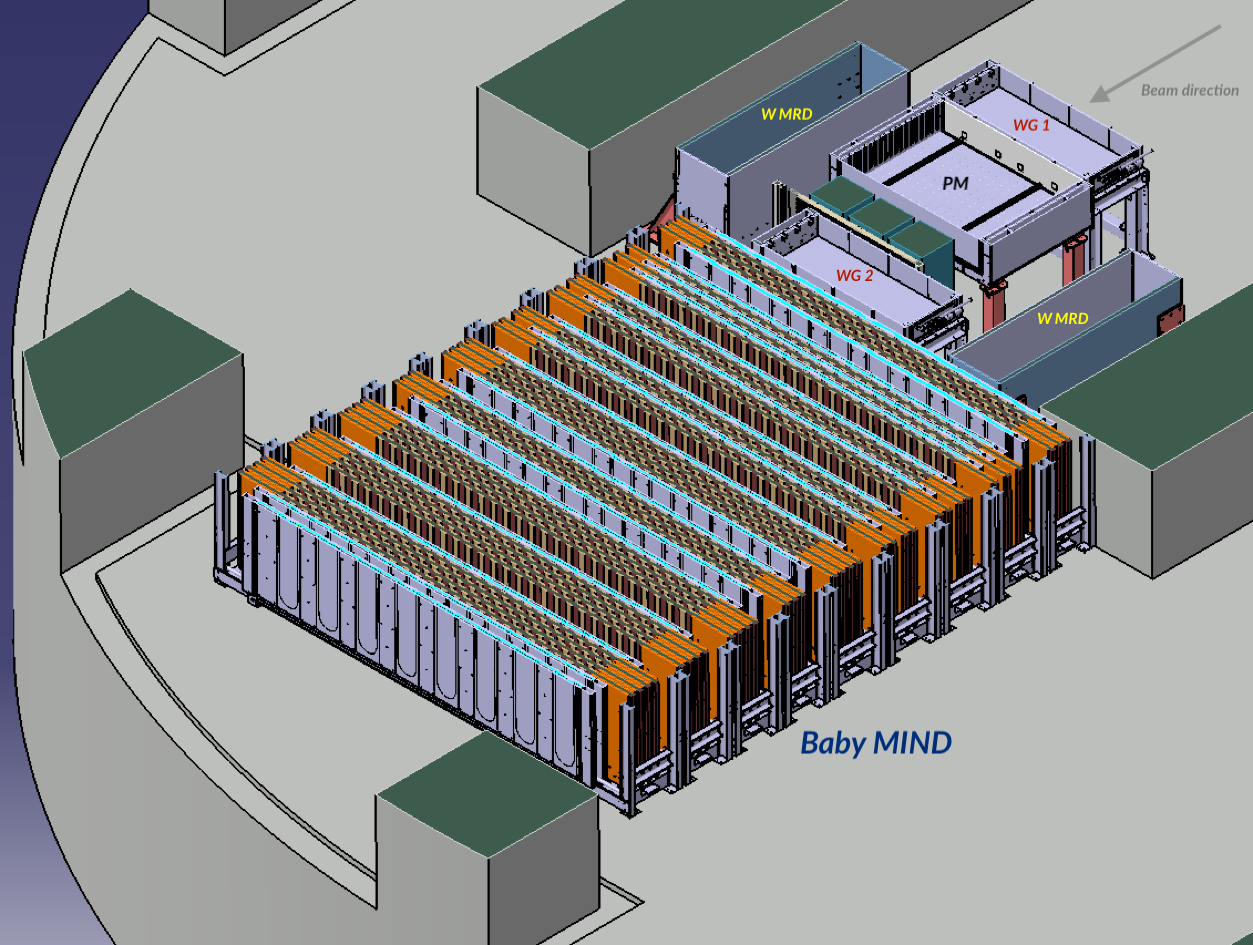}
	\caption{ Layout of WAGASCI sub-detectors (cross-sectional view).}\label{figLayout}
\end{minipage}%
\begin{minipage}{0.5\textwidth}
	\centering
	\includegraphics[height = 4cm]{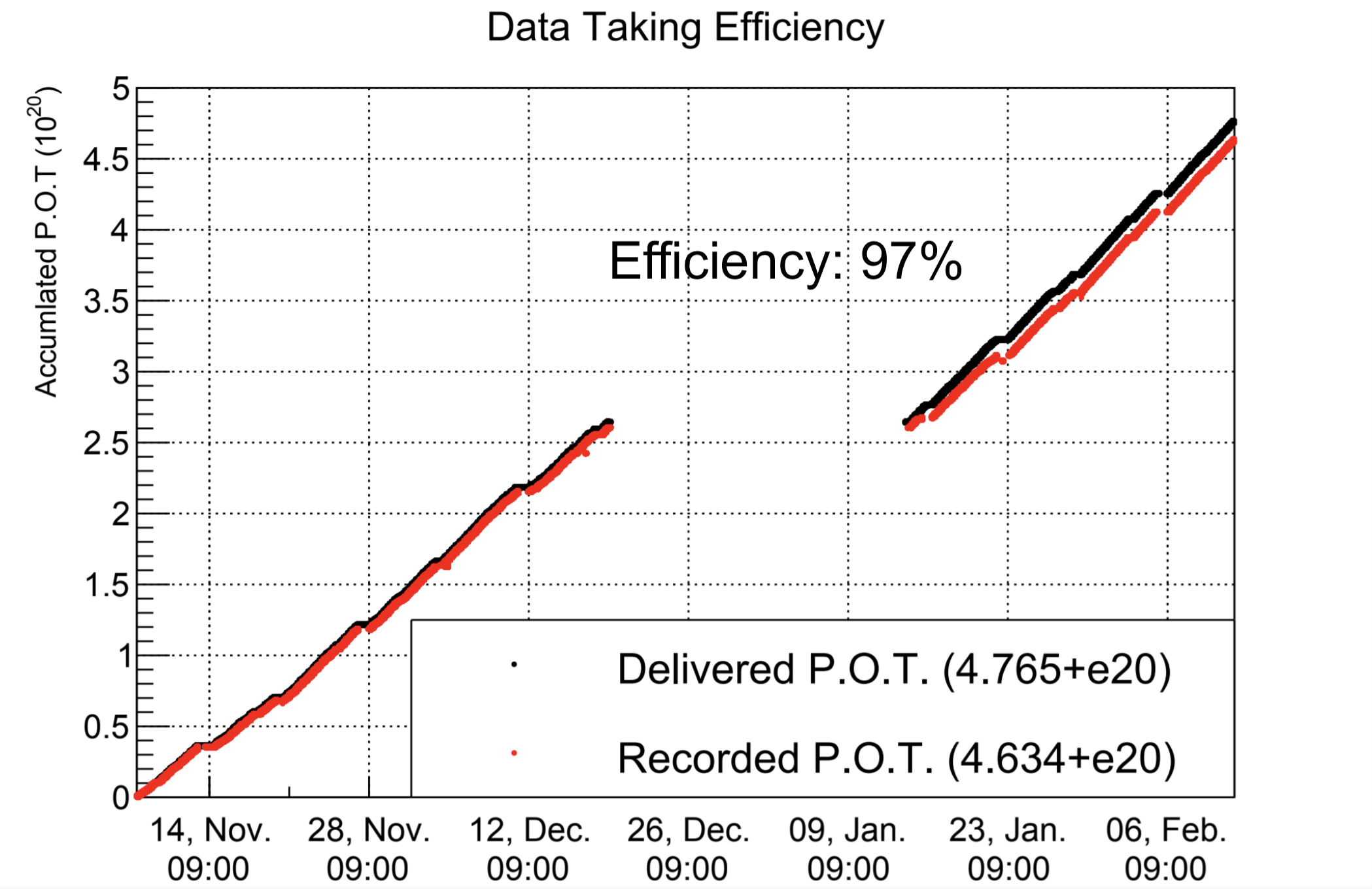}
	\caption{Baby MIND Data collection efficiency during Nov-Dec 2019.}\label{figEff}
\end{minipage}%
\end{figure}

Baby MIND \cite{babyMINDdesign} is the downstream muon range detector (MRD) of WAGASCI, located on the B2 floor of the T2K near detector complex, at a distance of $280~\rm{m}$ downstream from the T2K beam target. The configuration of all components of WAGASCI is shown in Figure \ref{figLayout}, including two WAGASCI target modules [WG], made of water and plastic, the Proton module [PM], made of plastic, two Wall MRDs [WMRD], made of iron and plastic and the Baby MIND detector. Detailed information about WAGASCI sub-detectors can be found in \cite{wagasci1}.

\noindent
While the main purpose of WAGASCI is to measure cross sections of charged current neutrino and antineutrino interactions on water and plastic, separate measurements can be obtained from the charged current neutrino and antineutrino interactions on iron in Baby MIND. With $33$ Magnet modules made of iron, each weighing $2$ tonnes, the expected number of neutrino interactions on iron in Baby MIND is far more significant than the WAGASCI target modules, which makes Baby MIND a suitable candidate for beam monitoring.  
The sensitive layers of the Baby MIND detector consist of $18$ scintillating modules, which are interleaved with magnet modules in an irregular pattern. Despite the low resolution of Baby MIND for tracking vertex activity, due to significant non-sensitive material in the particle's path, charged current neutrino interactions on iron with a muon traversing three or more scintillating modules can be reconstructed. A summary of Baby MIND's first physics run, the detector response and event pre-selection strategy are discussed in the following sections.

\section{Data collection efficiency during the first physics run}
The first physics run for WAGASCI-Baby MIND in $\nu_{\mu}$ beam mode took place between November 2019 and February 2020, coinciding with efforts to increase the T2K beam power, which eventually reached a record maximum power of  $522.6~\rm{kW}$. During this period (T2K run 10) a total of $4.8 \times 10^{20}$ Protons on target (POT) was delivered. Baby MIND succeeded in collecting the data in this period with $97~\%$ data collection efficiency (Figure \ref{figEff}), thanks to the stability of all detector systems including the magnet system, the electronics, the synchronization unit, the DAQ system and the efforts of the local operation team.

\section{Detector response}
The detector performance studies were carried out by selecting sand muons, which are the results of neutrino interactions in the material upstream of Baby MIND including other WAGASCI sub-detector. The requirement for having at least ten active modules in Baby MIND with low hit multiplicity aims to select clean high energy muons as highly penetrating Minimum Ionizing Particles (MIP).

\subsection{Readout and Calibration}
\begin{figure}[h!]
\centering
\begin{minipage}{0.5\textwidth}
	\centering
	\includegraphics[height = 3cm]{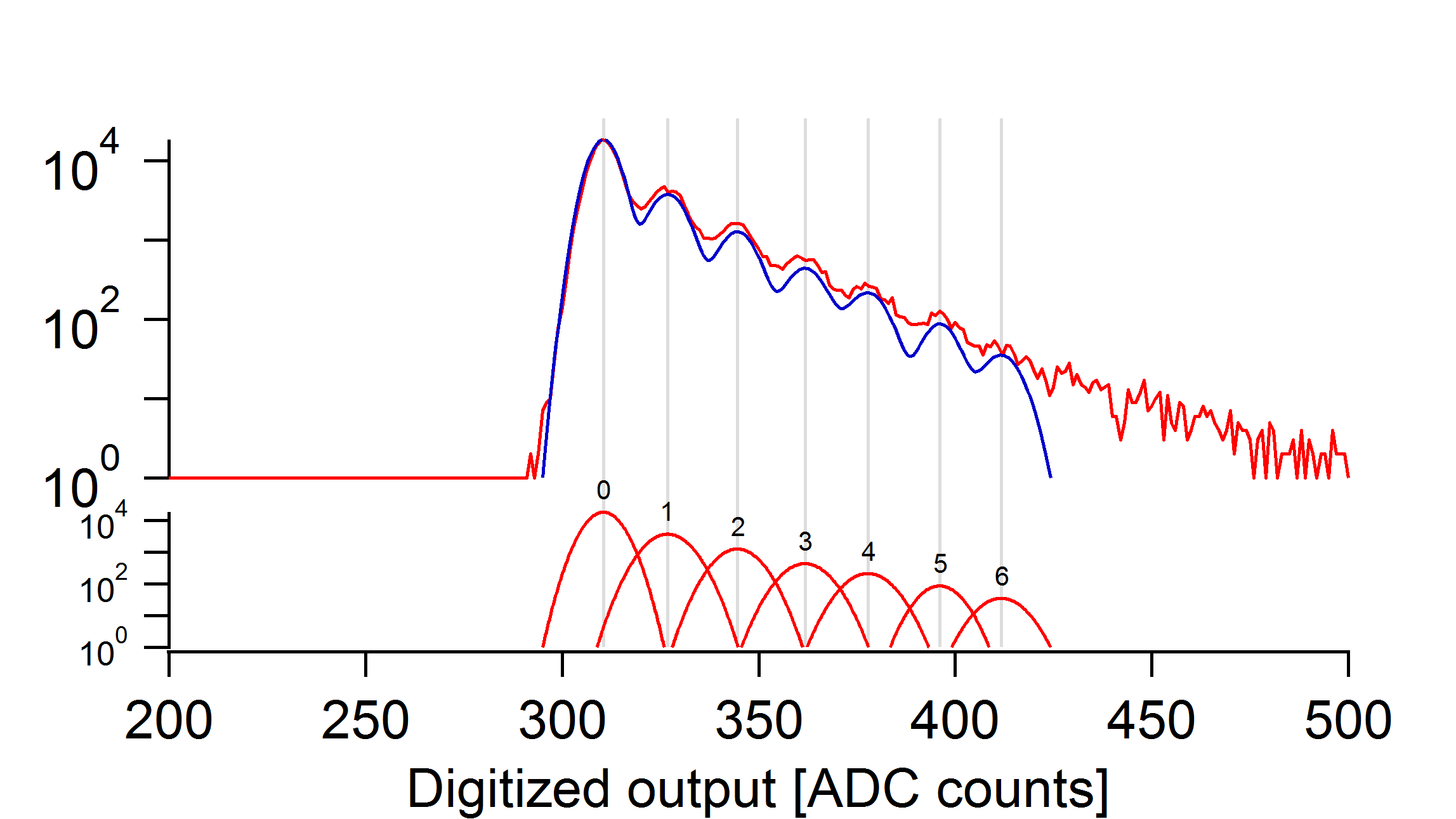}
	\caption{MPPC finger plot. Peaks represent 1, 2, 3, ... photo electrons respectively.}\label{figFingerplot}
\end{minipage}%
\begin{minipage}{0.5\textwidth}
	\centering
	\includegraphics[height = 3cm]{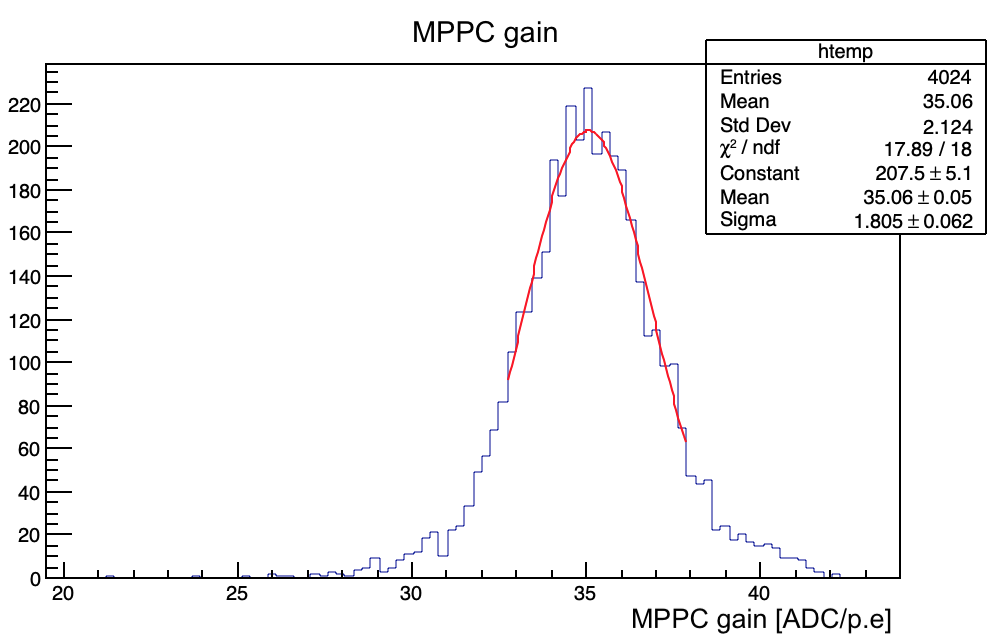}
	\caption{Distribution of MPPCs gain value [ADC/p.e].}\label{figgain}
	\end{minipage}%
\end{figure}

The Baby MIND readout scheme \cite{elec} consists of 3996 channels instrumented by Hamamatsu photosensors (Multi Pixel Photon Counter - MPPC) of type  S12571-025 and 28 channels by type S13081-050CS.
The readout electronics are based on the CITIROC chip. The MPPC input signal gets split in two paths, High Gain pre-amp (HG) and Low Gain pre-amp (LG), each with their own slow shapers and peak detection circuitry. The Time over Threshold (ToT) provides an additional measure of the signal strength. The HG analogue amplitude is calibrated by extracting the MPPC gain [ADC/p.e] from the dark count finger plot shown in Figures \ref{figFingerplot} and \ref{figgain}. HG is the most sensitive charge readout and has a range up to 100 p.e. The LG analogue amplitude is linearly proportional to HG amplitude before its saturation and has a larger range of up to 1000 p.e. Both of these analogue signal paths have a dead time of $10~\rm{{\mu}s}$ due to multiplexing and digitization which follows a tunable hold period designated for sampling the amplitudes. For this data run, the hold duration was set to $10~\mu\rm{s}$ to cover all of the eight bunches of a T2K spill. During a hold period only one value of HG and LG is registered per channel if above threshold. On the other hand, the digital signal path of ToT experiences no dead time but has a lower resolution. The calibration of ToT is achieved by fitting with respect to HG and LG. Finally the hit Light Yield (LY) is reconstructed by merging the information from the three signal paths.

\captionsetup[figure]{width = 0.4\linewidth}
\captionsetup[figure]{width = \linewidth}
\subsection{Light yield of scintillating bars}
\begin{figure}[h!]
\centering
\begin{minipage}{0.5\textwidth}
	\centering
	\includegraphics[height = 3.5cm]{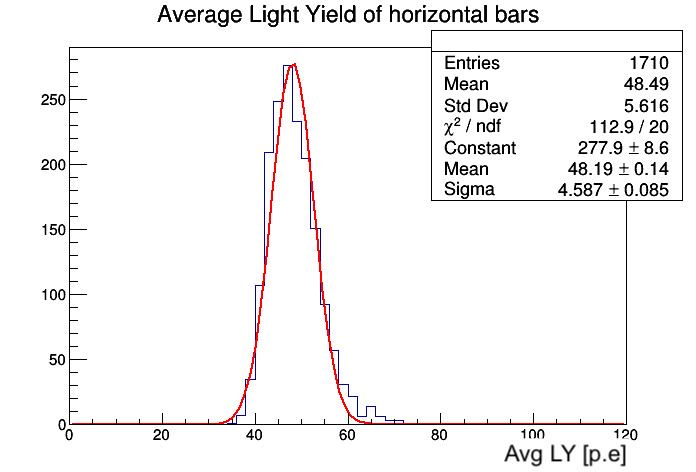}
\end{minipage}%
\begin{minipage}{0.5\textwidth}
	\centering
	\includegraphics[height = 3.5cm]{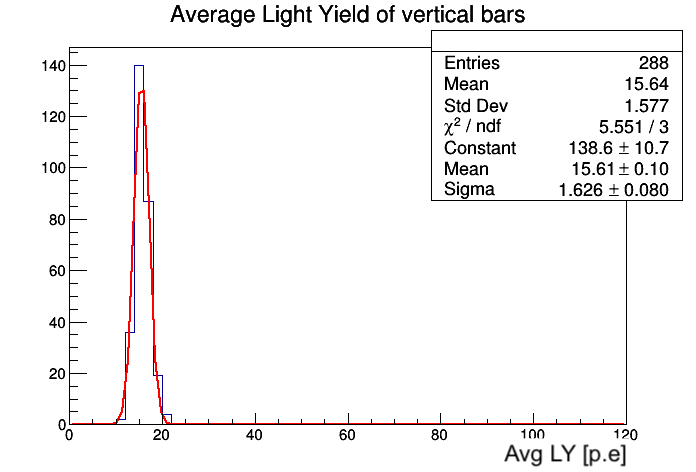}
	\end{minipage}%
	\caption{Distribution of the average LY summed over both ends for horizontal (left) and vertical bars (right).}\label{figLY}
\end{figure}

Each of the 18 scintillating modules contains 95 horizontal bars of $2900 \times 30 \times 7.5~\rm{mm}^3$ and 16 vertical bars of $1950 \times 210 \times 7.5~\rm{mm}^3$. The horizontal bars each has a straight wavelength shifting (WLS) fiber whose ends are read out from two sides, while the wider vertical bars each has a U shaped WLS fiber whose ends are both read out from the top.  Figure \ref{figLY} shows the distribution of the mean  LY, summed over both readout ends, for all horizontal (left) and vertical bars (right). 


\subsection{Timing}
\begin{figure}[h!]
\centering
\begin{minipage}{0.5\textwidth}
	\centering
	\includegraphics[height = 3.5cm]{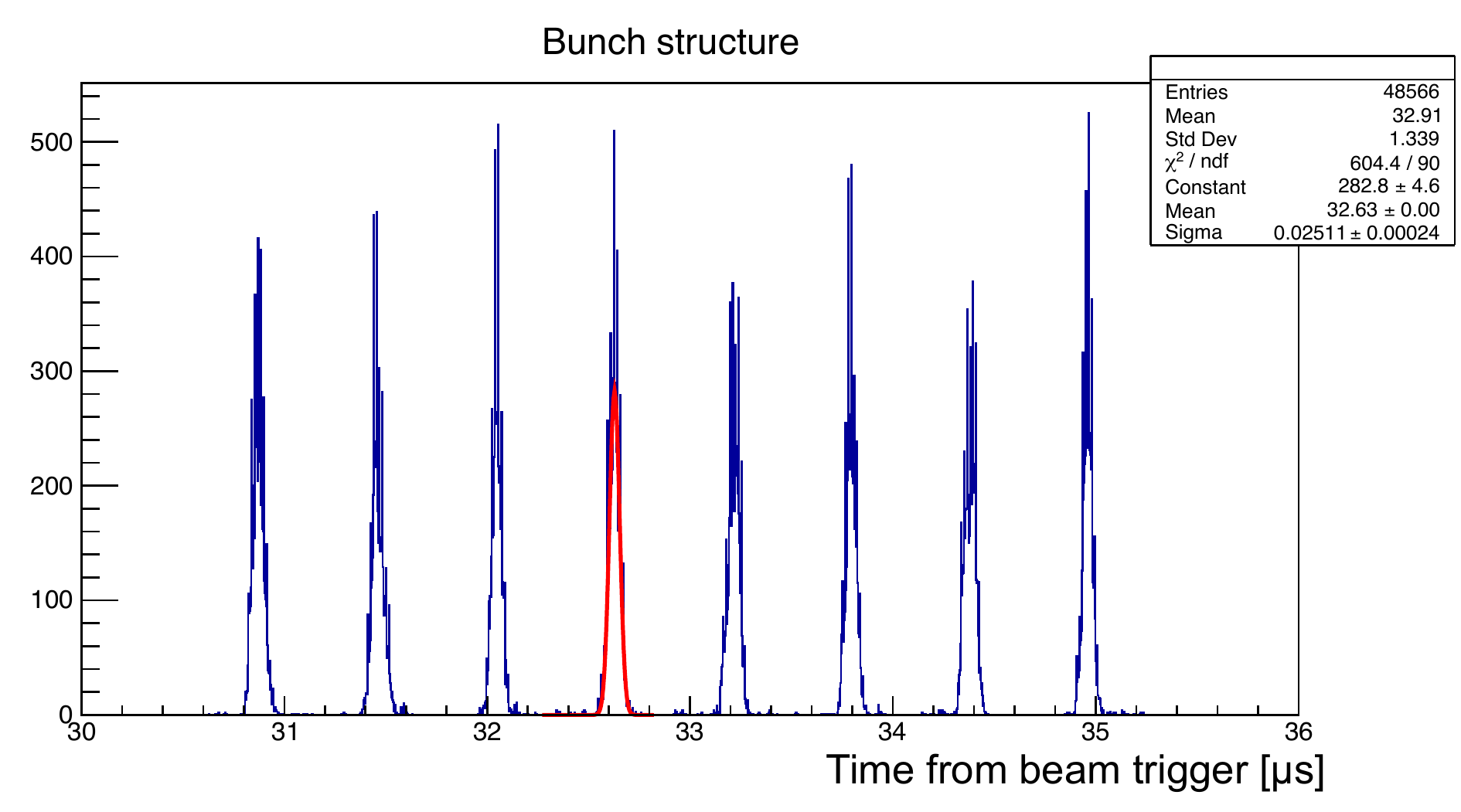}
	\caption{T2K bunch structure obtained from hit time distribution in Baby MIND.}\label{figBunch}
\end{minipage}%
\begin{minipage}{0.5\textwidth}
	\centering
	\includegraphics[height = 3.5cm]{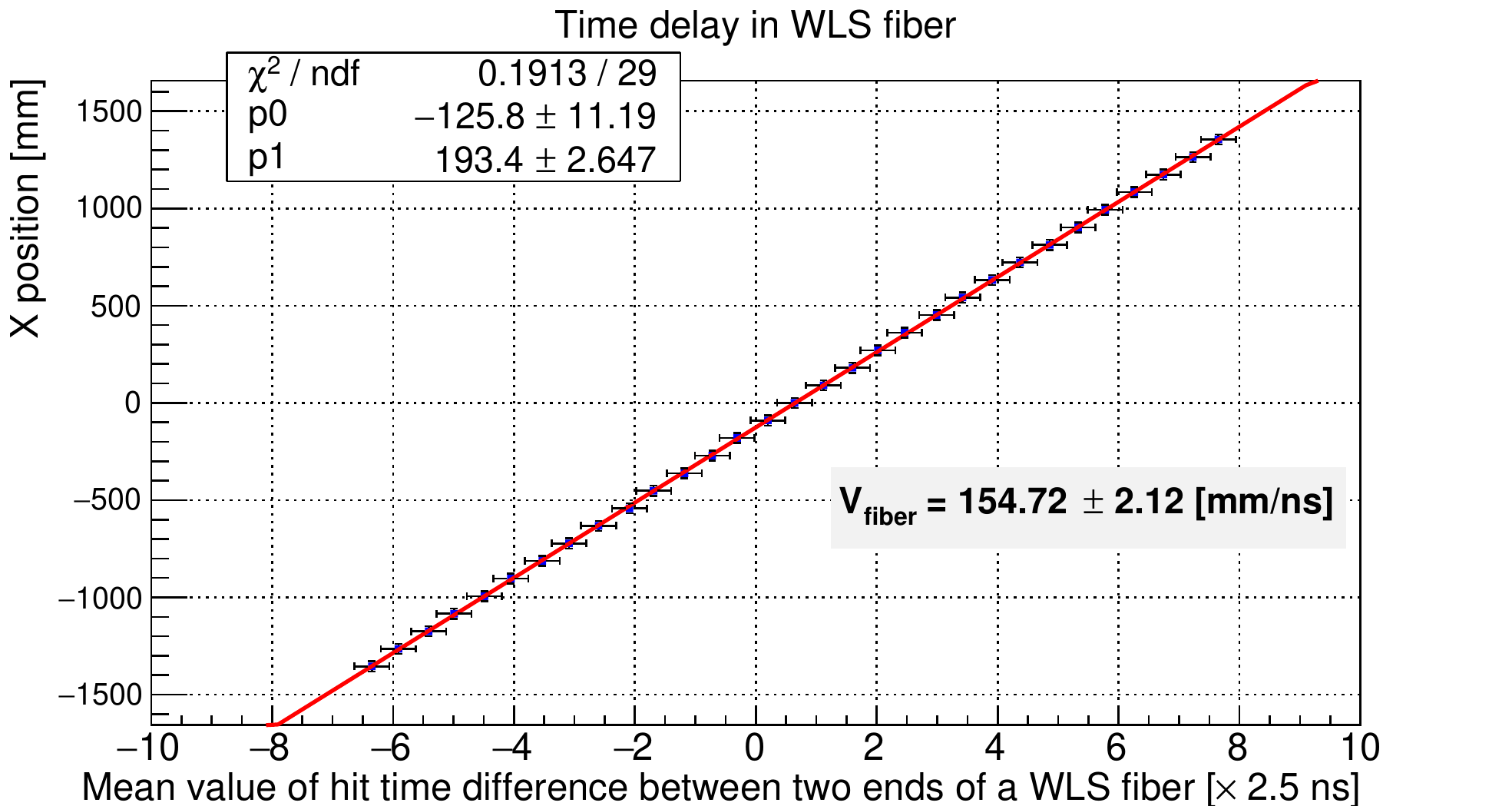}
	\caption{Velocity of propagation of light in WLS fiber.}\label{figFiberdelay}
	\end{minipage}%
\end{figure}

\begin{figure}[h!]
\centering
\begin{minipage}{0.5\textwidth}
	\centering
	\includegraphics[height = 3.5cm]{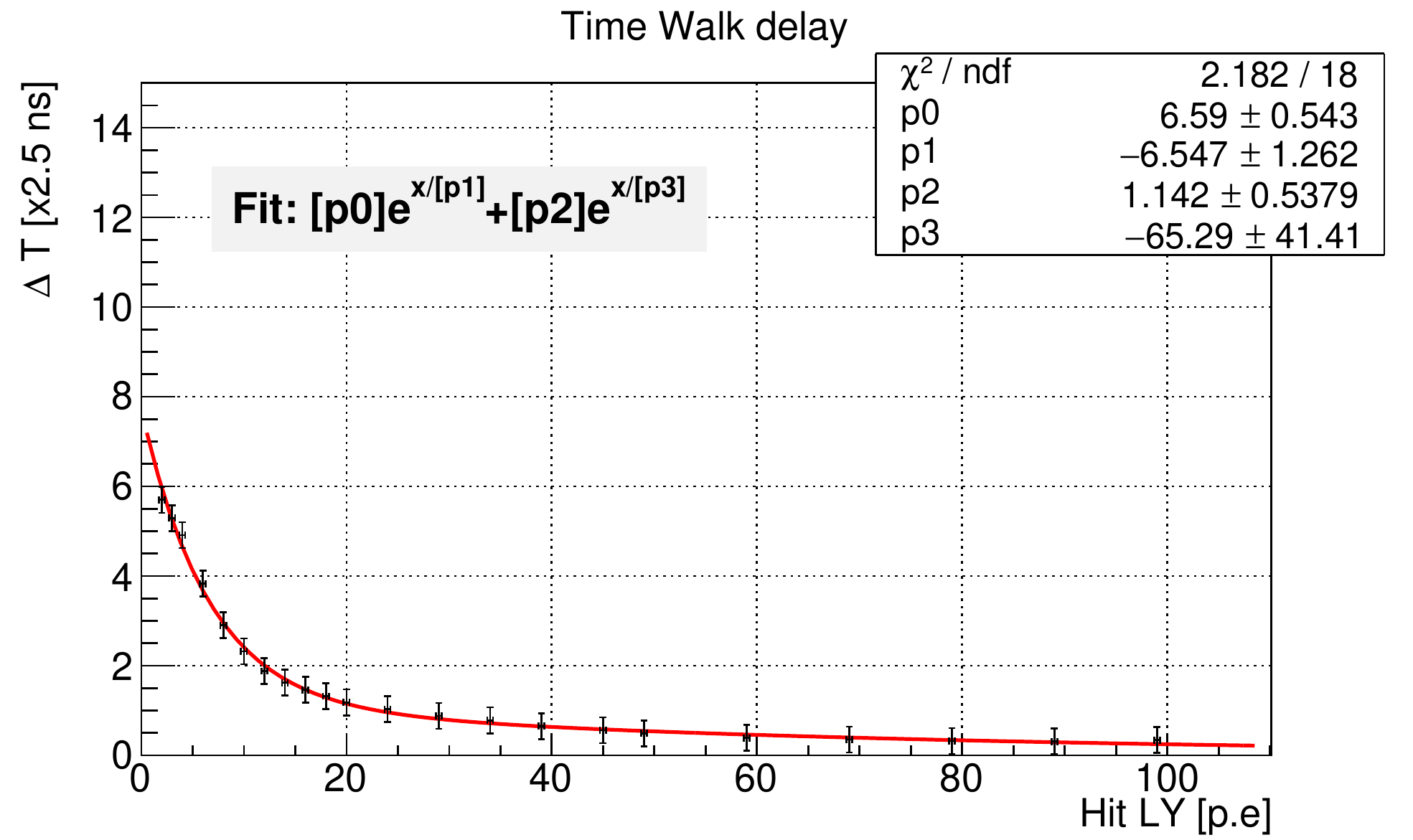}
	\caption{Time walk fit}\label{figTimewalk}
\end{minipage}%
\begin{minipage}{0.5\textwidth}
	\centering
	\includegraphics[height = 3.5cm]{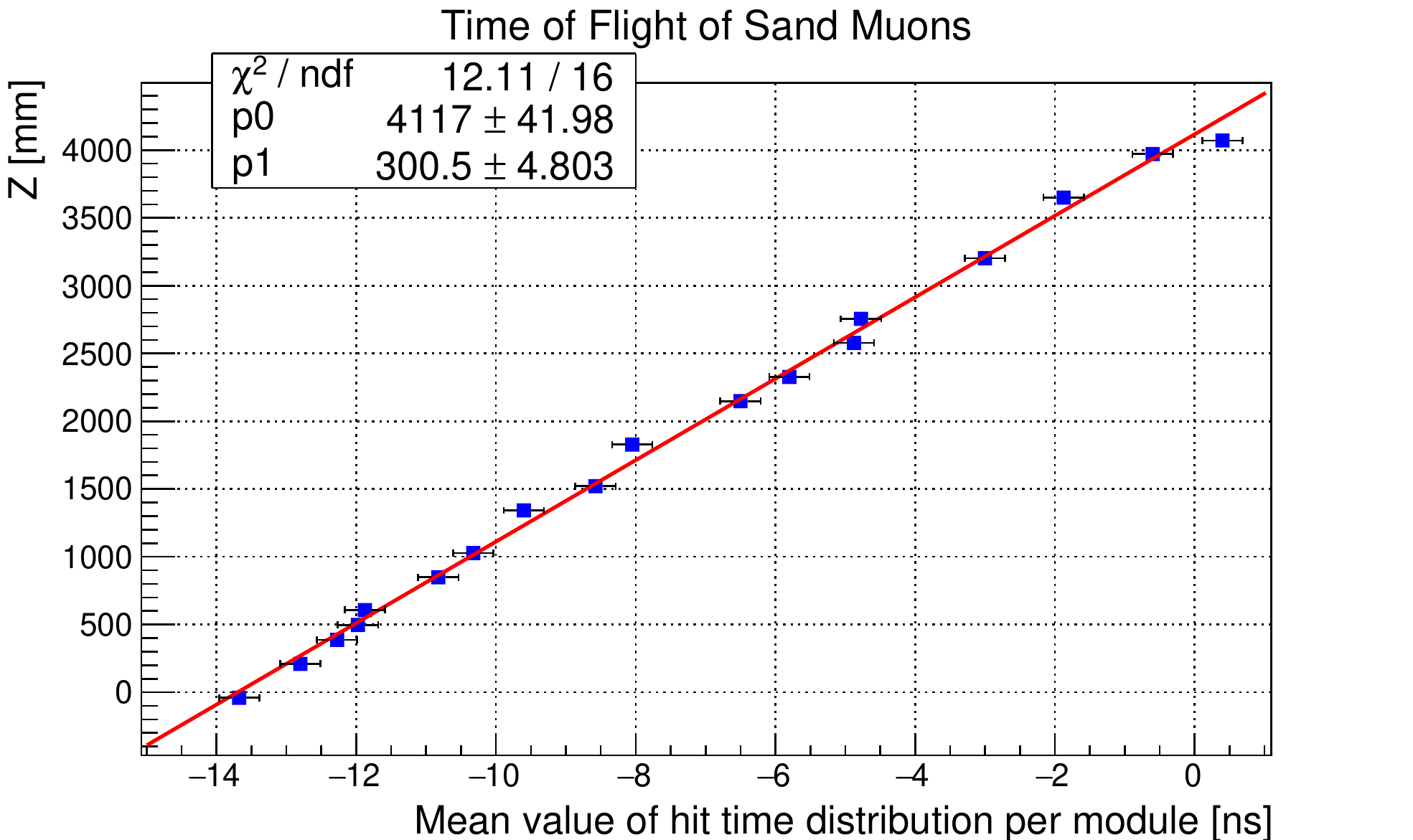}
	\caption{Time of flight of through-going sand muons.}\label{figTimeofFlight}
	\end{minipage}%
\end{figure}

The internal $400~\rm{MHz}$ clock of the Baby MIND electronics provides hit times in units of $2.5~\rm{ns}$. With this sampling frequency the bunch structure of T2K beam can be clearly seen in Figure \ref{figBunch}. The bunches are $580~\rm{ns}$ apart and have a sigma of $25~\rm{ns}$ each.
\captionsetup[figure]{width = 0.4\linewidth}
The hit time information is primarily used to group hits and pre-select events which occur in one bunch. In order to extract high accuracy timing information such as track directionality from the time of flight, some corrections need to be applied, namely the delays induced by WLS fiber and electronics time walk. Both studies have been carried out using the sand muons sample. The effective signal propagation velocity in WLS fiber of the horizontal bars was measured to be $154.72 \pm 2.12~\rm{mm/ns}$ (Figure \ref{figFiberdelay}) and the time walk was parametrized by a double exponential function as seen in Figure \ref{figTimewalk}. After these time corrections, the time of flight for through-going sand muons was investigated. In Figure \ref{figTimeofFlight}, each point represents the module $z$ position vs the mean value of the hit time distribution in the module relative to the hit time of the last module. Note that the angle and curvature of the tracks have not been corrected for. Based on this result, the direction of particles which have a track length of $1~\rm{m}$ or larger can be determined with the timing capabilities of Baby MIND.

\section{Event pre-selection}
The event pre-selection scheme can be described as follows. A pre-selected event is composed of two time separated objects (groups of hits) for side view (Y-Z) and top view (X-Z). Each view contains all the hits within $\pm 50 ~\rm{ns}$. The pre-selected events are categorized based on simple topology criteria as described in Table \ref{tab:categories}. The sand muons have been chosen to study the event rate stability over all data runs in the period of December 2019. The run statistics are summarized in Table \ref{tab:DecRuns}. The average occurrence rate of sand muons was found to be $0.92~\pm ~0.07$ events per $10^{15}$ POT in December data runs.
\begin{table}[h!]
\centering
\footnotesize
\caption{Event pre-selection categories.}
\begin{tabular}{@{}llp{6cm}@{}}
\hline
\hline
Categories & Description & Criteria \\
\hline
Pre type 0 & Short tracks & N active layers $ < 3$ (at least in one view)  \\

Pre type 1 & Clean single tracks &N active layers $ \geq 3$ (both views) \&\& \newline Cluster multiplicity per layer $\leq$ 2 \&\& \newline Cluster pos spread per layer $\leq 5$ cm\\

Pre type 2 & Other single tracks &  N active layers $ \geq 3$ (both views) \&\&\newline Cluster multiplicity per layer $\leq$ 4 \&\& \newline Cluster pos spread per layer $\leq 15$ cm\\

Pre type 3 & Multi tracks or pile up& all others events \\ 

\hline
&Sand Muons & (Pre type 1 $\parallel$ Pre type 2)\&\& \newline N active layers $\geq 10$ \&\& \newline  First layer is active\\
\hline
\hline
\end{tabular}  

\label{tab:categories}
 \end{table}
 
\begin{table}[h!]
\centering
\footnotesize
\caption{Summary of pre-selected events in December runs.}
\begin{tabular}{@{}lllllllp{1.8cm}@{}}
\hline
\hline
Date & N Spills  & Pre type 0 &Pre type 1 & Pre type 2 &Pre type 3& Sand Muons & SM Rate \tiny{ [/$10^{15}$ POT]}\\
\hline
1-Dec	&33248&	32664&	21985&	18329&	34676&	8120&	0.93	$\pm$0.06\\
2-Dec	&30295&	30490&	21044&	16620&	31844&	7531&	0.94	$\pm$0.07\\
3-Dec	&21697&	21626&	14830&	11730&	22387&	5376&	0.94	$\pm$0.08\\
4-Dec	&30384&	30250&	20412&	16301&	31336&	7229&	0.90	$\pm$0.07\\
5-Dec	&26089&	25783&	17464&	13991&	26813&	6446&	0.94	$\pm$0.07\\
6-Dec	&36417&	36541&	24826&	20291&	38937&	9004&	0.94	$\pm$0.06\\
7-Dec	&33581&	32707&	21761&	18735&	35138&	8115&	0.92	$\pm$0.06\\
8-Dec	&27831&	26853&	17897&	15190&	28457&	6583&	0.90	$\pm$0.07\\
9-Dec	&29015&	28882&	19090&	16127&	30770&	7139&	0.93	$\pm$0.07\\
10-Dec	&22336&	22016&	15104&	12077&	22914&	5313&	0.90	$\pm$0.08\\
12-Dec	&21204&	20865&	14249&	11189&	21835&	5113&	0.92	$\pm$0.08\\
13-Dec	&30605&	30172&	20059&	16810&	31922&	7463&	0.93	$\pm$0.07\\
14-Dec	&31613&	31780&	21245&	17428&	33365&	7683&	0.92	$\pm$0.06\\
15-Dec	&22984&	22466&	15137&	12652&	23831&	5598&	0.92	$\pm$0.08\\
16-Dec	&45086&	43748&	29243&	24129&	45381&	10711&	0.90	$\pm$0.05\\
17-Dec	&26569&	26274&	18096&	13997&	27297&	6455&	0.92	$\pm$0.07\\
\hline
\hline
\end{tabular}
\label{tab:DecRuns}
\end{table}

\captionsetup[figure]{width = \linewidth}
\begin{figure}[h!]
\centering
\includegraphics[width =8cm]{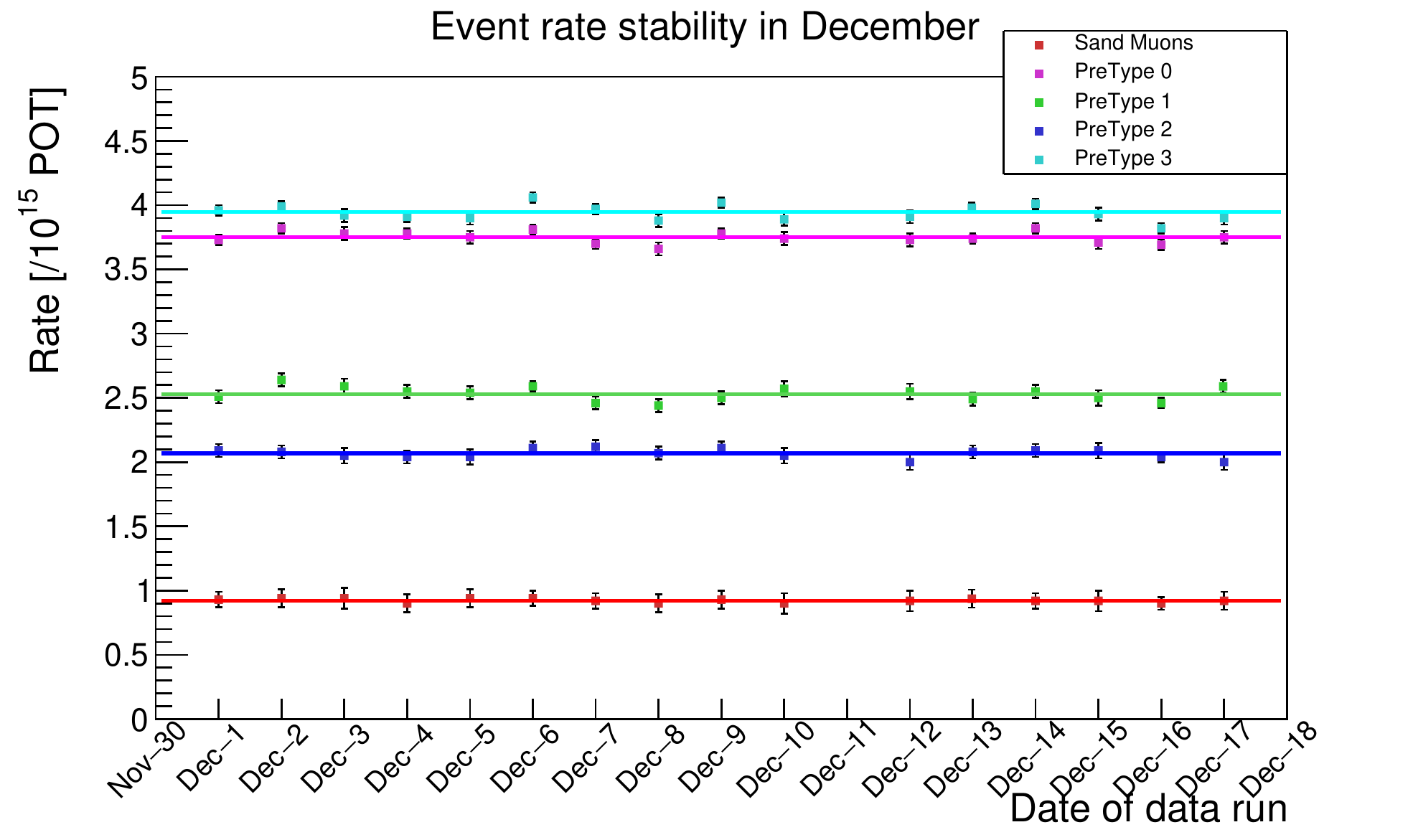}
\caption{Pre-selected event rate in December data runs. }\label{figEventRate}
\end{figure}

\section{Neutrino interactions on iron}
\begin{figure}[h!]
\centering
\begin{minipage}{0.5\textwidth}
	\centering
\includegraphics[height=3.8cm]{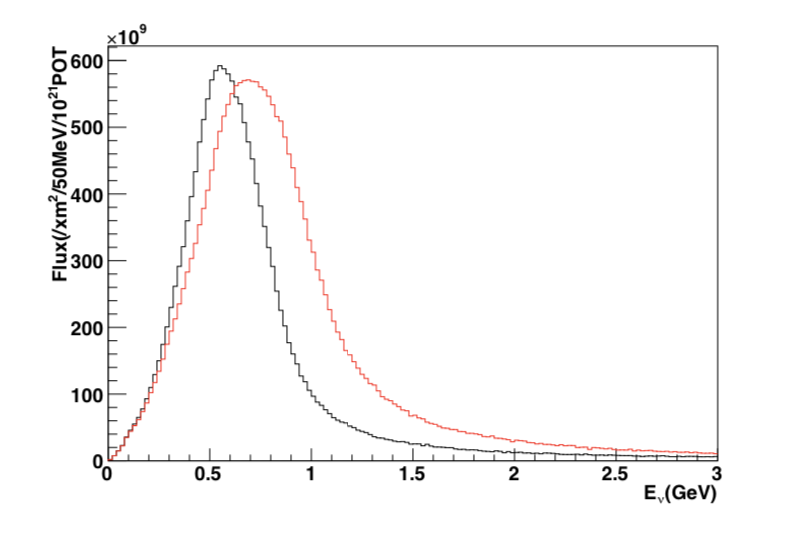}
\caption{Neutrino Energy spectru, at Baby MIND position (red, $1.6~^{\circ}$ off-axis) and at ND280 position (black, $2.5~^{\circ}$)}\label{figflux}
\end{minipage}%
\begin{minipage}{0.5\textwidth}
	\centering
	\includegraphics[height=3.8cm]{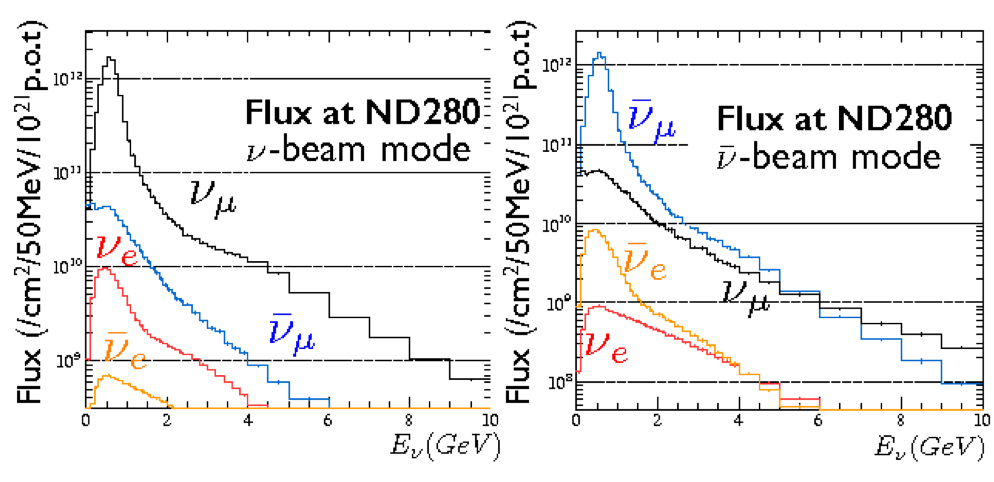}
\caption{T2K beam contamination with other neutrino types.}\label{figbackground}
	\end{minipage}%
\end{figure}

\begin{figure}[h!]
\centering
\begin{minipage}{0.5\textwidth}
	\centering
	\includegraphics[height = 2.6cm]{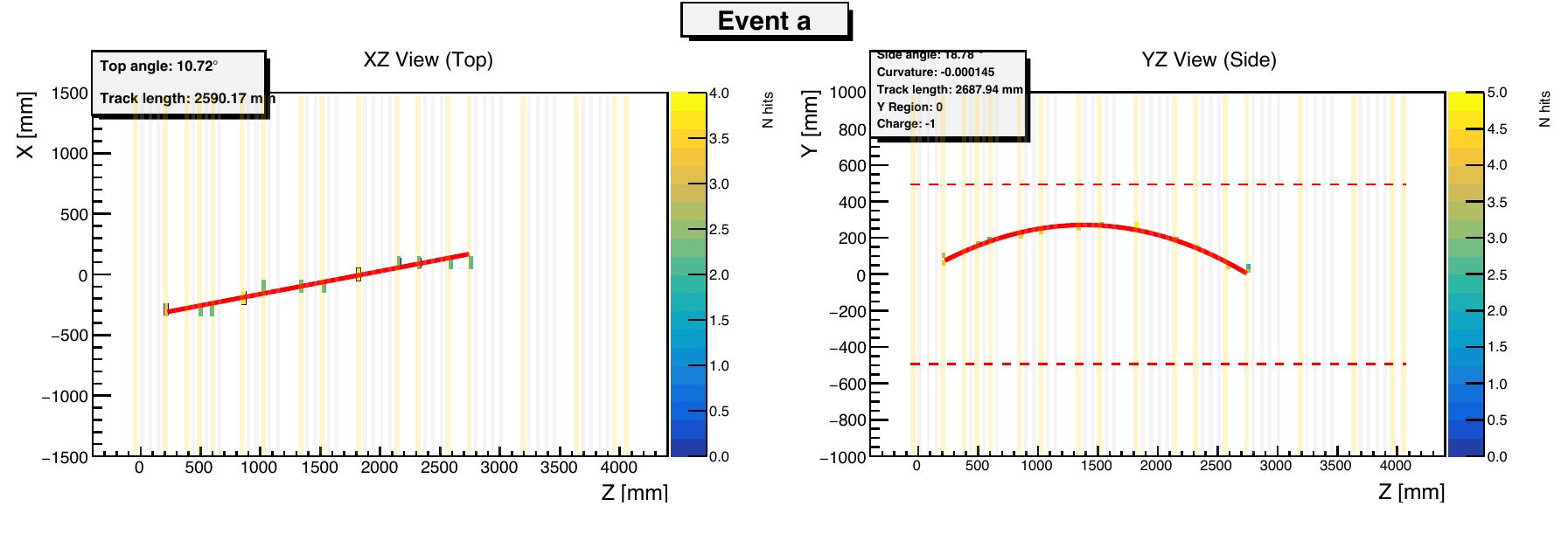}
\end{minipage}%
\begin{minipage}{0.5\textwidth}
	\centering
	\includegraphics[height = 2.8cm]{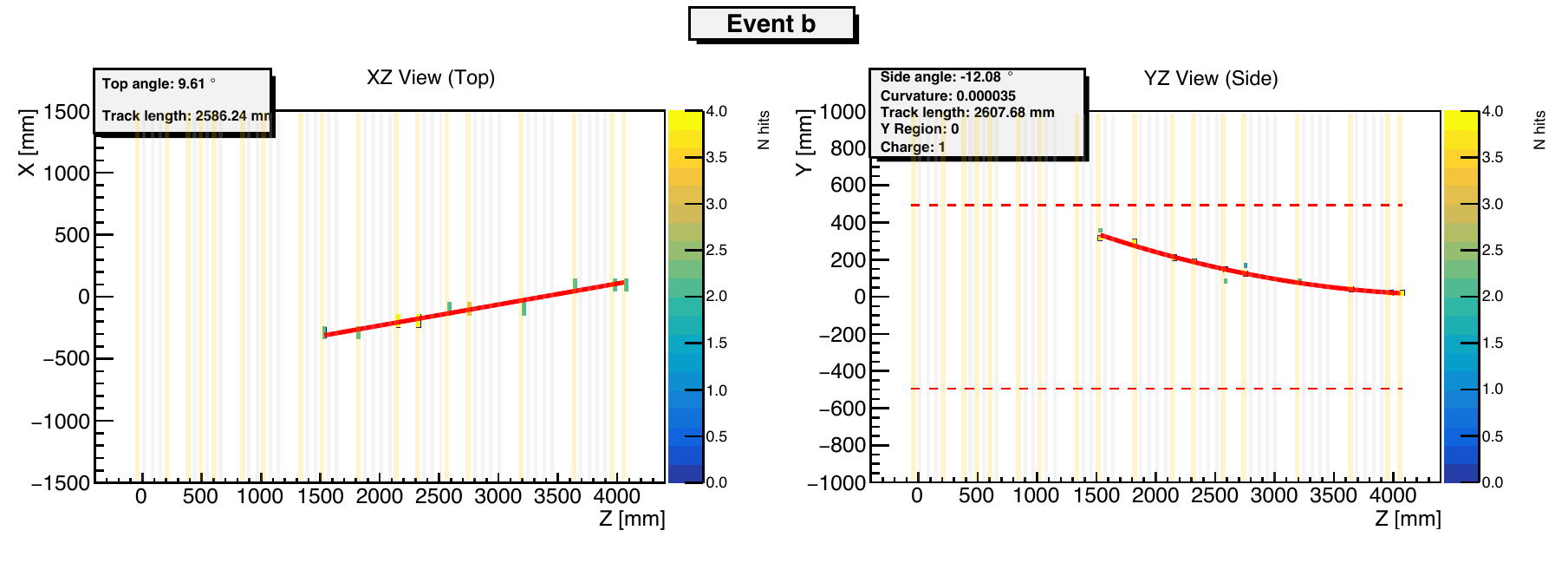}
\end{minipage}%
	\caption{Top view and side view of a $\nu_{\mu}$-CC-QE candidate with a $\mu^{-}$  bending downwards in the central magnet region (left), same for a $\overline{\nu}_{\mu}$-CC-QE candidate with a $_{•}{\mu^{+}}$ bending upwards (right).}\label{figmucandidates}
\end{figure}

At an off-axis angle of $1.6~^{\circ}$ from the T2K beam direction, Baby MIND receives neutrinos with slightly higher energies than ND280 (Figure \ref{figflux}). The T2K $\nu_{\mu}$ and $\overline{\nu}_{\mu}$ beams have a small fraction of wrong sign background (Figure \ref{figbackground}), which can generate a muon of opposite charge in case of a charge current interaction. Baby MIND can be used to monitor the fraction of the wrong sign background with a high charge ID efficiency. Figure \ref{figmucandidates} shows such candidate events from the data. The tracks are bent in the $1.5~\rm{T}$ magnetic field of the Baby MIND magnet modules \cite{magnet}. 
\captionsetup[figure]{width = 0.4\linewidth}

\begin{figure}[h!]
\centering
\begin{minipage}{0.5\textwidth}
	\centering
	\includegraphics[height = 2.5cm]{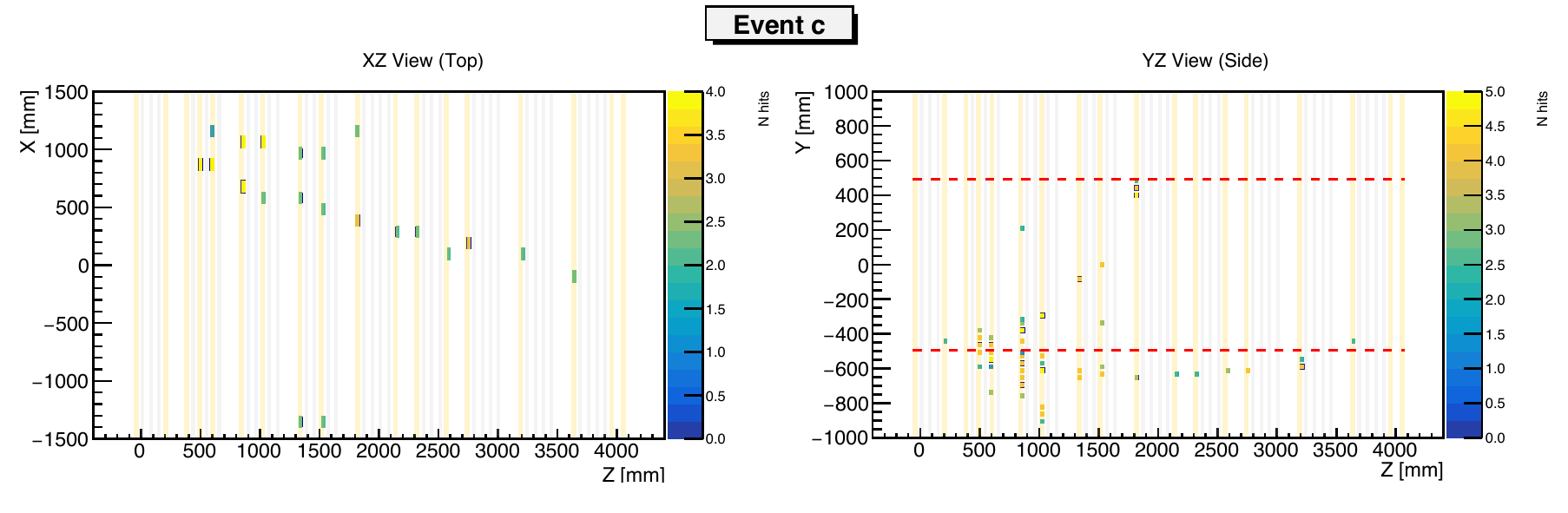}
\end{minipage}%
\begin{minipage}{0.5\textwidth}
	\centering
	\includegraphics[height = 2.7cm]{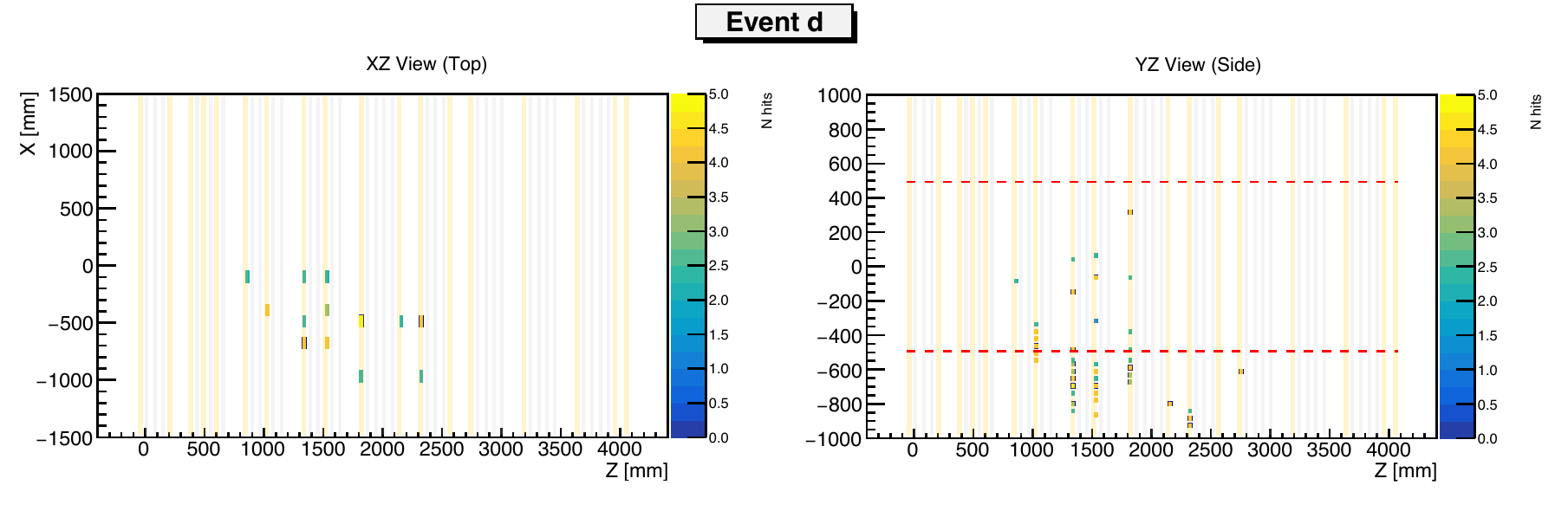}
\end{minipage}%
	\caption{Top view and side view of two events with multi track and shower like topologies.}\label{figothercandidates}
\end{figure}

\captionsetup[figure]{width =\linewidth}
\noindent However, Baby MIND observes many neutrino interactions that are not as clean and easy to reconstruct. Monte Carlo (MC) studies with neutrino flux and energies expected at the location of Baby MIND, can reveal the expected contributions of different $\nu_{\mu}$ interaction modes on iron, which is summarized in Table \ref{tab1}.  Examples of events with multi track or shower like topologies are presented in Figure \ref{figothercandidates}.
\begin{table}[h!]
\small
\begin{center}
\caption{Contributions of different $\nu_{\mu}$ interaction modes on iron (from Monte Carlo).} \label{tab1}
\begin{tabular}{cllll}
\hline
\hline
Interaction mode & CC-QE & CC-$1\pi$ & CC-n$\pi$ & NC \\
\hline
$\%$ & $37.6$ & $21$ & $15.5$ & $26$  \\
\hline
\hline
\end{tabular}
\end{center}
\end{table}

\captionsetup[figure]{width =\linewidth}
%
\section{Summary}
The first physics run of Baby MIND together with other WAGASCI subdetectors has been a successful data taking campaign with $97~\%$ data collection efficiency. The detector performance studies such as light yield, timing and event rate stability studies have been reported and examples of neutrino interactions on iron have been presented.   Prospects for the use of Baby MIND as a magnetized beam monitor for the T2K beam, with the capability of measuring the wrong sign component of the (anti)neutrino beam is under investigation. \\
 


\acknowledgement{This project has received funding from the European Union’s Horizon 2020 Research and Innovation programme under grant agreement No. 654168.}

\end{document}